\documentclass[english,usenatbib]{mn2e}
\usepackage{times}
\usepackage{array}
\usepackage{calc}
\usepackage{graphicx}
\IfFileExists{url.sty}{\usepackage{url}}
                      {\newcommand{\url}{\texttt}}
\usepackage[authoryear]{natbib}

\makeatletter

\newcommand{\noun}[1]{\textsc{#1}}

\providecommand{\tabularnewline}{\\}

\usepackage{aas_macros}
\usepackage{url}

\usepackage{babel}
\makeatother

\begin{document}
\title[Rapidly processing masks of next-gen surveys]{Methods for rapidly processing angular masks of next-generation galaxy surveys}
\date{Accepted 2008 April 5. Received 2008 March 7; in original form 2007 December 2}
\author[M. E. C. Swanson et al.]{Molly E. C. Swanson,$^{1}$\thanks{E-mail: molly@space.mit.edu} Max Tegmark,$^{1}$ Andrew J. S. Hamilton,$^{2}$ and J. Colin Hill$^{1}$\\  $^{1}$Dept.~of Physics and MIT Kavli Institute, Massachusetts Institute of Technology, 77 Massachusetts Ave, Cambridge, MA 02139,USA\\  $^{2}$JILA and Dept.~of Astrophysical and Planetary Sciences, Box 440, University of Colorado, Boulder, CO 80309, USA} 
\maketitle

\begin{abstract}
As galaxy surveys become larger and more complex, keeping track of
the completeness, magnitude limit, and other survey parameters as
a function of direction on the sky becomes an increasingly challenging
computational task. For example, typical angular masks of the Sloan
Digital Sky Survey contain about $N=300{,}000$ distinct spherical
polygons. Managing masks with such large numbers of polygons becomes
intractably slow, particularly for tasks that run in time $\mathcal{O}\left(N^{2}\right)$
with a naive algorithm, such as finding which polygons overlap each
other. Here we present a {}``divide-and-conquer'' solution to this
challenge: we first split the angular mask into predefined regions
called {}``pixels,'' such that each polygon is in only one pixel,
and then perform further computations, such as checking for overlap,
on the polygons within each pixel separately. This reduces $\mathcal{O}\left(N^{2}\right)$
tasks to $\mathcal{O}\left(N\right)$, and also reduces the important
task of determining in which polygon(s) a point on the sky lies from
$\mathcal{O}\left(N\right)$ to $\mathcal{O}\left(1\right)$, resulting
in significant computational speedup. Additionally, we present a method
to efficiently convert any angular mask to and from the popular \emph{HEALPix}
format. This method can be generically applied to convert to and from
any desired spherical pixelization. We have implemented these techniques
in a new version of the \noun{mangle} software package, which is freely
available at \texttt{\url{http://space.mit.edu/home/tegmark/mangle/}},
along with complete documentation and example applications. These
new methods should prove quite useful to the astronomical community,
and since \noun{mangle} is a generic tool for managing angular masks
on a sphere, it has the potential to benefit terrestrial mapmaking
applications as well.
\end{abstract}

\section{Introduction}

\begin{figure*}
\includegraphics{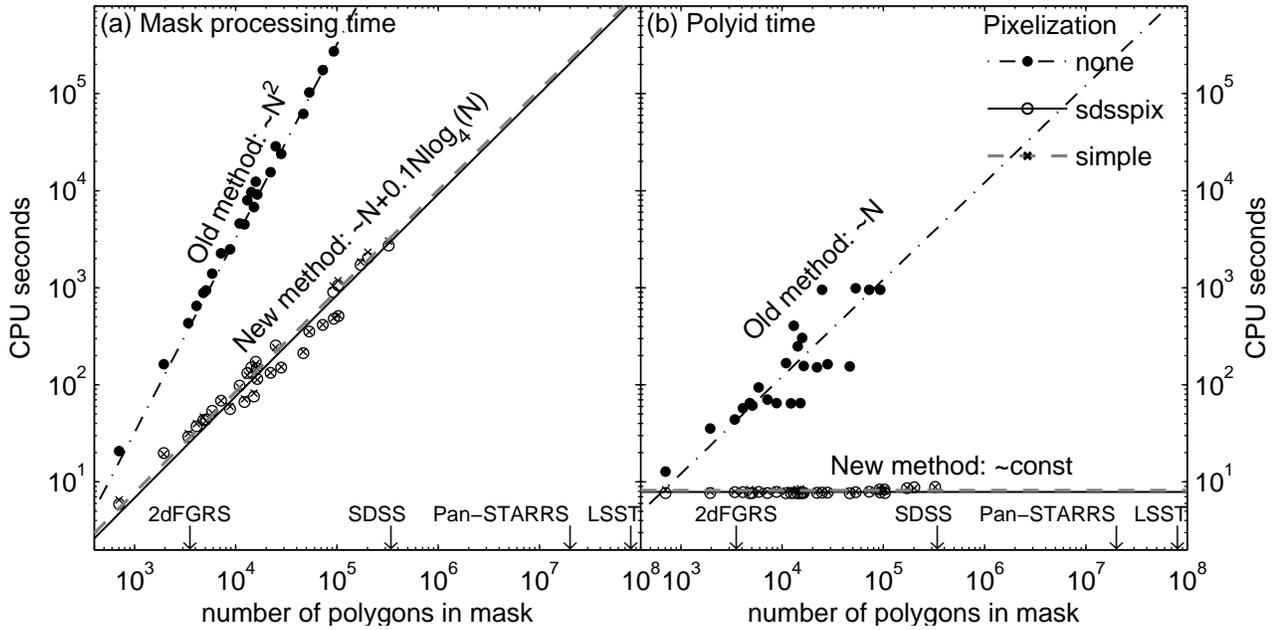}

\caption{\label{fig:speed_fractions}Speed trials for a series of portions
of the SDSS DR5 mask with and without pixelization. (a) Time required
for pixelization, snapping, balkanization and unification of the mask.
(b) Time required to identify in which polygon each of the ${\sim}400{,}000$
SDSS DR5 galaxies lies. Each set of trials is fitted with a power
law to show how the processing time scales with the number of polygons
$N$. Also shown on the x-axis are the number of polygons in the 2dFGRS
mask, the SDSS DR5 mask, and conservative estimates for the Pan-STARRS
and LSST large-scale structure masks based on scaling up 2dFGRS.}
\end{figure*}
Over the past few decades, galaxy surveys have provided a wealth of
information about the large-scale structure of our Universe, and the
next generation of surveys currently being planned promises to provide
even more insight. In order to realize the full potential of upcoming
surveys, it is essential to avoid unnecessary errors and approximations
in the way they are analysed. The tremendous volumes of data produced
by these new surveys will shrink statistical uncertainty to unprecedented
levels, and in order to take advantage of this we must ensure that
the systematic uncertainties can keep pace. The purpose of this paper
is to maximize the scientific utility of next-generation surveys by
providing methods for processing angular masks as rapidly and accurately
as possible.

Angular masks of a galaxy survey are functions of direction on the
sky that model the survey completeness, magnitude limit, seeing, dust
extinction, or other parameters that vary across the sky. The earliest
galaxy redshift surveys -- the first Center for Astrophysics redshift
survey (CfA1; \citealt{1983ApJS...52...89H}) and the first Southern
Sky Redshift Survey (SSRS1; \citealt{1991ApJS...75..935D}) -- had
simple angular masks defined by boundaries in declination and Galactic
latitude. The next generation of surveys -- IRAS \citep{1992ApJS...83...29S}
and PCSz \citep{2000MNRAS.317...55S} -- had somewhat more complex
masks, with some regions of high contamination excluded from the survey.

The present generation of surveys -- the Two Degree Field Galaxy Redshift
Survey (2dFGRS; \citealt{2001MNRAS.328.1039C,2003astro.ph..6581C})
and the Sloan Digital Sky Survey (SDSS; \citealt{2000AJ....120.1579Y})
-- consist of photometric surveys that identify galaxies and measure
their angular positions combined with spectroscopic surveys that measure
a redshift for each galaxy to determine its distance from us. Angular
masks are useful for describing parameters for both photometric and
spectroscopic surveys -- for example, seeing and magnitude limit are
key parameters to model in photometric surveys, and the survey completeness
-- i.e., the fraction of photometrically selected target galaxies
for which a spectrum has been measured -- is vital for analysing spectroscopic
surveys.

The angular masks of SDSS and 2dFGRS consist of circular fields defined
by the spectroscopic plates of the redshift survey superimposed on
an angular mask of the parent photometric survey. 2dFGRS uses the
Automatic Plate Measurement (APM) survey (Maddox et al. 1990a,b, 1996)
\nocite{1990MNRAS.243..692M,1990MNRAS.246..433M,1996MNRAS.283.1227M}
as its parent photometric survey and covers approximately 1500 square
degrees. The APM angular mask consists of 269 $5^{\circ}\times5^{\circ}$
photographic plates which are drilled with numerous holes to avoid
bright stars, satellite trails, plate defects, and so forth. Combining
the photographic plates, holes, and spectroscopic fields gives a total
of 3525 polygons that define the spectroscopic angular mask of 2dFGRS.

The SDSS covers a larger area on the sky -- 5740 square degrees in
Data Release 5 (DR\%; \citealt{2007ApJS..172..634A}) -- and has a
yet more complicated angular mask than 2dFGRS. The SDSS photometric
survey is done by drift-scanning: each scan across the sky covers
six long, narrow scanlines, and the gaps between these lines are filled
in with a second scan slightly offset from the first, producing a
{}``stripe'' about $2.5^{\circ}$ wide assembled from 12 scanlines.
In addition to the fairly intricate pattern produced by this scanning
strategy, there are nearly 250,000 holes masked out of the photometric
survey for various reasons, plus the circular $3^{\circ}$ spectroscopic
fields. Combining all of these elements produces an angular mask for
the spectroscopic survey that contains 340,351 polygons.

To accurately manage the 2dFGRS and SDSS angular masks, \citet{2004MNRAS.349..115H}
developed a suite of general-purpose software called \noun{mangle},
which performs several important procedures on angular masks using
computational methods detailed in \citet{1993ApJ...406L..47H,1993ApJ...417...19H}.
This software has proved to be a valuable resource to the astronomical
community: it has been used in several analyses of galaxy survey data
\citep{2002MNRAS.335..887T,2004ApJ...606..702T,2005MNRAS.361.1287M,2005PASJ...57..709H,2005ApJ...633...11P,2005ApJ...635..990C,2007ApJ...658..898P,2007PASJ...59...93N,2007AJ....133.2222S,2007ApJ...664..608W,2007arXiv0707.3445T}.
Additionally, it was used extensively in the preparation of the New
York University Value-Added Galaxy Catalog (NYU-VAGC; \citealt{2005AJ....129.2562B}),
which has been used as the basis for almost all publications on large-scale
structure by the SDSS collaboration.

However, many of functions in the original version of \noun{mangle}
run in $\mathcal{O}\left(N^{2}\right)$ time, which becomes quite
computationally challenging as the size and complexity of surveys
continues to increase -- computations involving the SDSS mask can
take several months of CPU time. In this paper we present new algorithms
that can process complicated angular masks such as SDSS dramatically
faster with no loss of accuracy. Our method is based on splitting
an angular mask into pixels, reducing the processing time to $\mathcal{O}\left(N\right)$
by adding an $\mathcal{O}\left(N\log N\right)$ preprocessing step. 
Similar methods based on hierarchical spatial subdivisions have been
found to be useful in the field of computational geometry
(see, e.g., \citealt{bigbluebook}),
but have not previously been applied to angular masks in an astronomy context.

This will be especially useful for next-generation surveys, such as
the Dark Energy Survey (DES; \citealt{2005astro.ph.10346T}) and surveys
done with the Wide-Field Multi-Object Spectrograph (WFMOS; \citealt{2006PhRvD..74f3525Y,wfmos}),
the Panoramic Survey Telescope and Rapid Response System (Pan-STARRS;
\citealt{2004SPIE.5489...11K}), and the Large Synoptic Survey Telescope
(LSST; \citealt{2002SPIE.4836...10T,2004AAS...20510802S,2006AIPC..870...44T,LSST}).
DES, Pan-STARRS, and LSST will perform photometric surveys and use
techniques for estimating redshifts based on the photometric information;
WFMOS will perform spectroscopic surveys using one of the upcoming
photometric surveys for target selection. The methods we present here
are useful for both photometric and spectroscopic surveys -- keeping
track of factors such as seeing and dust extinction could prove particularly
important for photometric redshift determinations \citep{2007MNRAS.375...68C,2008ApJ...674..768O,2008MNRAS.386.1219B}.

\begin{figure*}
\includegraphics{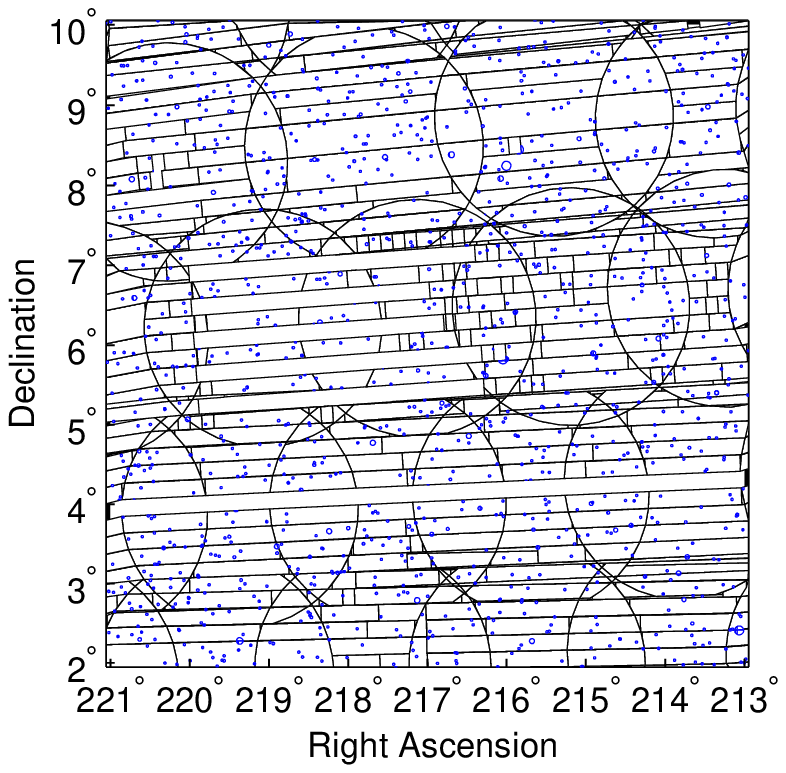}\includegraphics{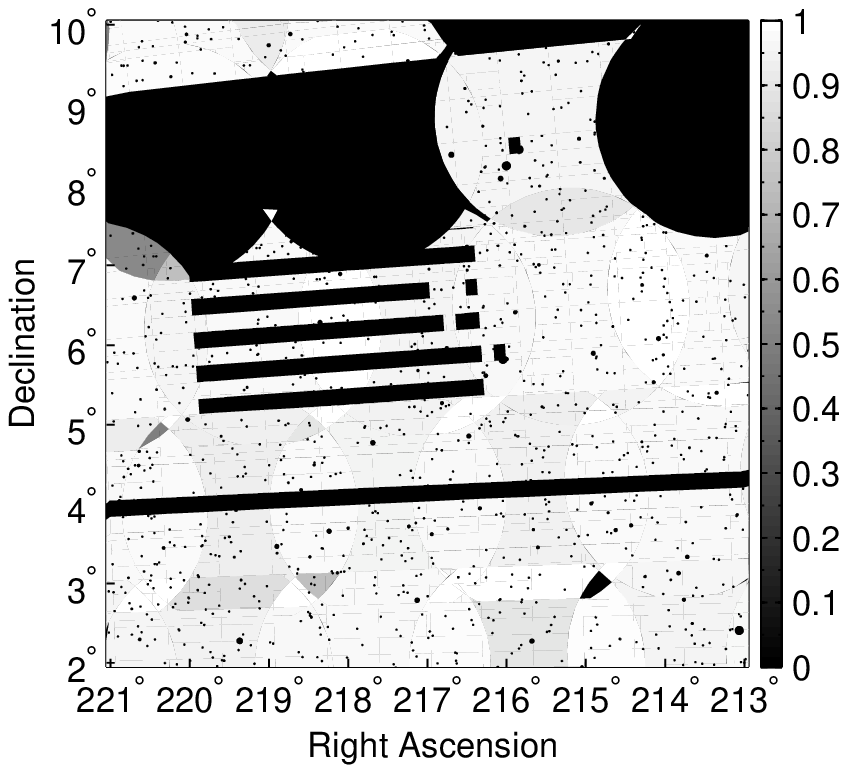}

\caption{\label{fig:examplemask}\label{fig:examplemask_final}A portion of
the SDSS DR5 angular mask \citep{2005AJ....129.2562B,2007ApJS..172..634A}.
Left: Polygons defining the mask: spectroscopic plates and lines delineating
different scans and spectroscopic plates are shown in black, and holes
in the mask are shown in blue/gray. Right: Processed version of the
mask, shaded according to survey completeness.}
\end{figure*}
\begin{table*}
\begin{minipage}[c][70mm]{175mm}%

\caption{\label{tab:mangleterms}Definitions of Terms, in Alphabetical Order}

\begin{tabular}{@{}lp{155mm}}
Term &
Definition \tabularnewline
\hline
boundary &
A set of edges bounding a polygon. \tabularnewline
cap &
A spherical disk, a region above a circle on the unit sphere. \tabularnewline
circle &
A line of constant latitude with respect to some arbitrary polar axis
on the unit sphere. \tabularnewline
edge &
An edge is part of a circle. A polygon is enclosed by its edges. \tabularnewline
great circle &
A line of zero latitude with respect to some arbitrary polar axis
on the unit sphere. A great circle is a circle, but a circle is not
necessarily a great circle. \tabularnewline
mask &
The union of an arbitrary number of weighted polygons. \tabularnewline
pixel&
A special polygon that specifies some predefined region on the sky
as part of a scheme for discretizing the unit sphere. Once a mask
has been pixelized, each polygon in the mask is guaranteed to overlap
with exactly one pixel.\tabularnewline
polygon &
The intersection of an arbitrary number of caps. \tabularnewline
rectangle &
A special kind of polygon, a rectangular polygon bounded by lines
of constant longitude and latitude. \tabularnewline
vertex &
A point of intersection of two circles. A vertex of a polygon is a
point where two of its edges meet. \tabularnewline
weight &
The weight assigned to a polygon. The spherical harmonics of a mask
are the sum of the spherical harmonics of its polygons, each weighted
according to its weight. A weight of 1 is the usual weight. A weight
of 0 signifies an empty polygon, a hole. In general the weight may
be some arbitrary positive or negative real number. \tabularnewline
\hline
\end{tabular}%
\end{minipage}%

\end{table*}
The proposed large-scale structure survey to be produced by Pan-STARRS
will cover ${\sim}30{,}000$ square degrees in 5 wavelength bands,
with each field being observed ${\sim}50$ times such that the images
can be co-added. A naive scaling up of the 2dFGRS area and number
of polygons gives an estimate of ${\sim}2\times10^7$ polygons for
the final Pan-STARRS mask. Similarly, the LSST large-scale structure
survey will cover ${\sim}20{,}000$ square degrees in 6 bands, with
${\sim}200$ co-added images, suggesting ${\sim}8\times10^7$ polygons
for the LSST mask. The need for an improvement in mask processing
speed is clearly illustrated in Fig.~\ref{fig:speed_fractions}:
with the old algorithms, the projected processing time for the LSST
mask would be over 6000 years. With our new method, this time is reduced
by a factor of ${\sim}24{,}000$ to just ten days.

\begin{figure*}
\includegraphics{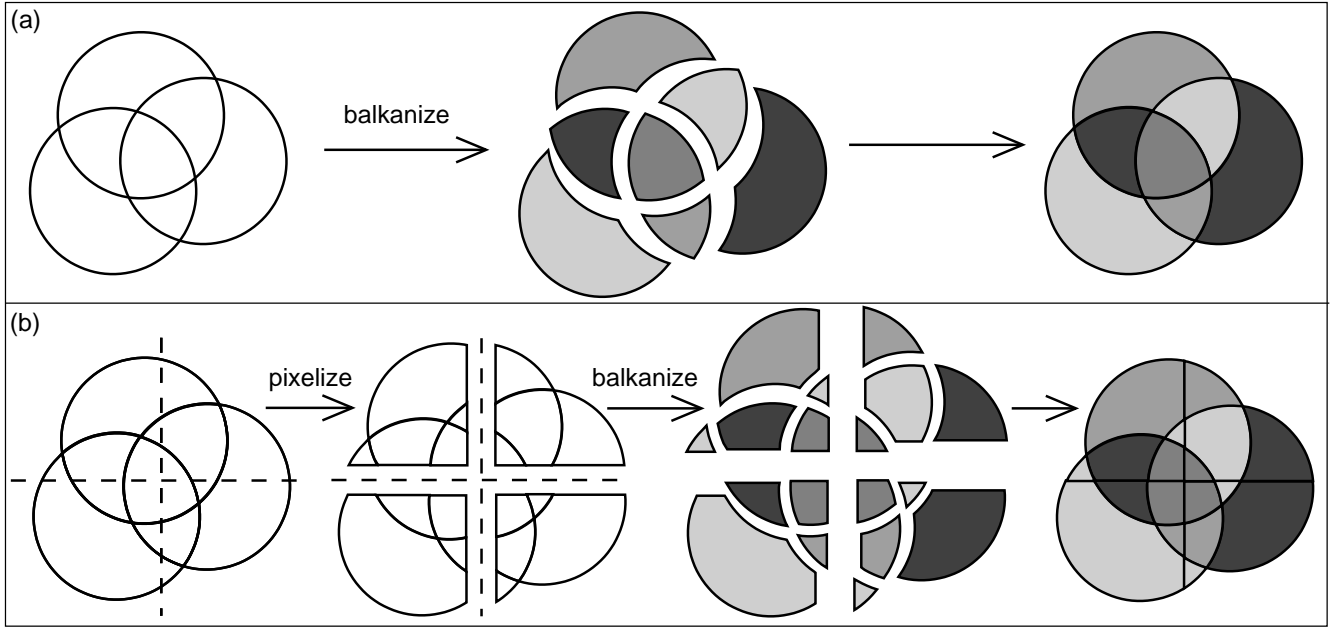}

\caption{\label{fig:pix_concept}A cartoon illustrating the process of balkanization
(a) with no pixelization, (b) with pixelization.}
\end{figure*}
In addition to developing faster algorithms for processing angular
masks, we have also integrated the \noun{mangle} utilities with \emph{HEALPix},
a widely used tool for discretizing the celestial sphere \citep{2005ApJ...622..759G}.
The methods used by \noun{mangle} are complementary to \emph{HEALPix}:
\noun{mangle} is best used for functions that are piecewise-constant
in distinct regions of the sky, such as the completeness of a galaxy
survey. In contrast, \emph{HEALPix} is optimal for describing functions
that are continuously varying across the sky, such as the cosmic microwave
background (CMB) or the amount of extinction due to Galactic dust.
The ability to convert rapidly between these two formats allows for
easy comparison of these two types of data without the unnecessary
approximation inherent in discretizing an angular mask. Furthermore,
converting a mask into \emph{HEALPix} format allows users to take
advantage of pre-existing \emph{HEALPix} tools for rapidly computing
spherical harmonics. 

The spectacular surveys on the horizon are preparing to generate massive,
powerful datasets that will be made publically available -- this in
turn necessitates powerful and intuitive general-purpose tools that
assist the community to do science with this avalanche of data. We
provide a such tools with this new generation of the \noun{mangle}
software and describe these new tools here. However, this paper is
not a software manual (a manual is provided on the \noun{mangle} website)
but rather a description of the underlying algorithms. These tools
have been utilized in recent analyses of SDSS data \citep{2006PhRvD..74l3507T,2008MNRAS.385.1635S};
we are now making them public so others can use them as well.

The outline of this paper is as follows: in \S\ref{sec:Mangle-terminology}
we give an overview of the terminology we use to describe angular
masks and the basic tasks we wish to perform, and in \S\ref{sec:Speedup:-pixelization}
we detail our algorithms for accelerating these tasks and quantify
their speed. We describe our methods for integrating \noun{mangle}
with \emph{HEALPix} in \S\ref{sec:Unification-with-HEALPix} and
summarize in \S\ref{sec:Summary}.


\section{\noun{Mangle} terminology}

\label{sec:Mangle-terminology}

The process of defining an angular mask of a galaxy survey in a generic
way requires a set of standardized terminology. We use the terminology
from \citet{2004MNRAS.349..115H} and present a summary of it here.
Our formal definition of an angular mask is a union of an arbitrary
number of weighted angular regions bounded by arbitrary numbers of
edges. The restrictions on the mask are

\begin{enumerate}
\item that each edge must be part of some circle on the sphere (but not
necessarily a great circle), and
\item that the weight within each subregion of the mask must be constant.
\end{enumerate}
This definition does not cover every theoretical possibility of how
a piecewise-constant function on a sphere could be defined, but in
practice it is sufficiently broad to accommodate the design of essentially
any galaxy survey. Furthermore, as we discuss in detail in \S\ref{sec:Unification-with-HEALPix},
a curvilinear angular region (such as a \emph{HEALPix} pixel) can
be well-approximated by segments of circles at high resolution. As
an example of a typical angular mask, we show a portion of the SDSS
angular mask from Data Release 5 (DR5; \citealt{2007ApJS..172..634A})
in Fig.~\ref{fig:examplemask}. 

The fundamental building block of an angular mask is the spherical
polygon, which is defined as a region bounded by edges that are part
of a circle on the sphere. An angular mask is thus the union of arbitrarily
weighted non-overlapping polygons. For convenience, we provide an
updated version of a table from \citet{2004MNRAS.349..115H} in Table~\ref{tab:mangleterms}
containing the definitions of key terms used in this paper. 

The basic procedure used by the \noun{mangle} software to process
an angular mask consists of the following steps, which are described
in greater detail in \citet{2004MNRAS.349..115H}:

\begin{enumerate}
\item Snap
\item Balkanize
\item Weight
\item Unify
\end{enumerate}
The snapping step identifies edges of polygons that are nearly coincident
and snaps them together so the edges line up exactly. This is necessary
because nearly-coincident edges can cause significant numerical issues
in later computations. In practice, this situation occurs when two
polygons in a survey are intended to abut perfectly, but are prevented
from doing so by roundoff errors or numerical imprecision in the mask
definition. There are several tunable tolerances that control how
close two edges can be before they get snapped together -- these can
be adjusted based on how precisely the polygons defining the mask
are specified.

Balkanization is the process of resolving a mask into a set of non-overlapping
polygons. It checks for overlap between each pair of polygons in the
mask, and if the polygons overlap, it fragments them into non-overlapping
pieces. After this is completed, it identifies polygons having disconnected
geometry and subdivides them into connected parts. The basic concept
of balkanization is illustrated in the top panel of Fig.~\ref{fig:pix_concept}.
The purpose of this procedure is to define all of the distinct regions
on the sky in which the piecewise-constant function we are intending
to model might take on a different value. For example, the 2dFGRS
and SDSS spectroscopic surveys generate masks containing many overlapping
circles defining each spectroscopic field observed. The SDSS spectrographs
can observe 640 objects in each field, so if there are more than 640
desired target galaxies in the field, they might not all be observed.
For example, one field may have spectra for 80 per cent of the targets,
and a neighbouring field may have 90 per cent, but in the region where
they overlap all of the targets may have been observed. This is how
the survey completeness is determined and illustrates why balkanization
is necessary.

After the mask is balkanized, weights are assigned to each polygon,
representing the value of the survey completeness (or any other desired
parameter) in that region. The way this is done depends on how this
information is provided for a given survey. For example, the 2dFGRS
mask software by Peder Norberg and Shaun Cole\footnote{\url{http://magnum.anu.edu.au/~TDFgg/Public/Release/Masks}}
provides a function that takes an angular position on the sky and
returns the completeness, the magnitude limit, the photographic plate
number, and the value of the parameter $\mu$ (described in \citealt{2001MNRAS.328.1039C})
at that location. This information can be imported into \noun{mangle}
by producing a list of the midpoints of each of the polygons in the
mask, applying the 2dFGRS software to calculate the value of the desired
function at each midpoint, and assigning these weights to the appropriate
polygon by using the {}``weight'' routine in \noun{mangle}. For
the SDSS mask, the files provided by the NYU-VAGC\footnote{\url{http://sdss.physics.nyu.edu/vagc}}
\citep{2005AJ....129.2562B} already include weights for the survey
completeness, so this step is not needed.

The final step of the processing is unification, which discards polygons
with zero weight and combines neighbouring polygons that have the
same weight. While not strictly necessary, this procedure clears out
unneeded clutter and makes subsequent calculations more efficient.

After an angular mask has been processed in this fashion, it can be
used for function evaluation: i.e., given a point on the sky, determine
in which one of the non-overlapping polygons it lies, and then get
the weight of that polygon to obtain the value of the function at
the input point. It can also be used for creating a random sample
of points with the same selection function as the survey, calculating
Data-Random $\left\langle DR\right\rangle $ and Random-Random $\left\langle RR\right\rangle $
angular integrals, and computing the spherical harmonics of the mask.
The \noun{mangle} software provides utilities for all of these tasks.

\section{Speedup: pixelization}

\label{sec:Speedup:-pixelization}

The tasks of snapping, balkanization, and unification all require
comparing pairs of polygons in the mask. The brute-force method to
accomplish this is simply to compare each polygon with every other
polygon, which is what the original version of \noun{mangle} did.
This naive algorithm is $\mathcal{O}\left(N^{2}\right)$, which is
easily sufficient for masks such as the 2dFGRS mask, with an $N$
of a few thousand polygons. However, the SDSS mask has about 100 times
as many polygons, and points to the need for a cleverer approach.
The method we present here is a divide-and-conquer approach we dub
{}``pixelization,'' which processes the mask so that each polygon
needs to be compared with only a few nearby polygons.

\subsection{Pixelization concept}

\label{sub:Pixelization-concept}

\begin{figure*}
\includegraphics{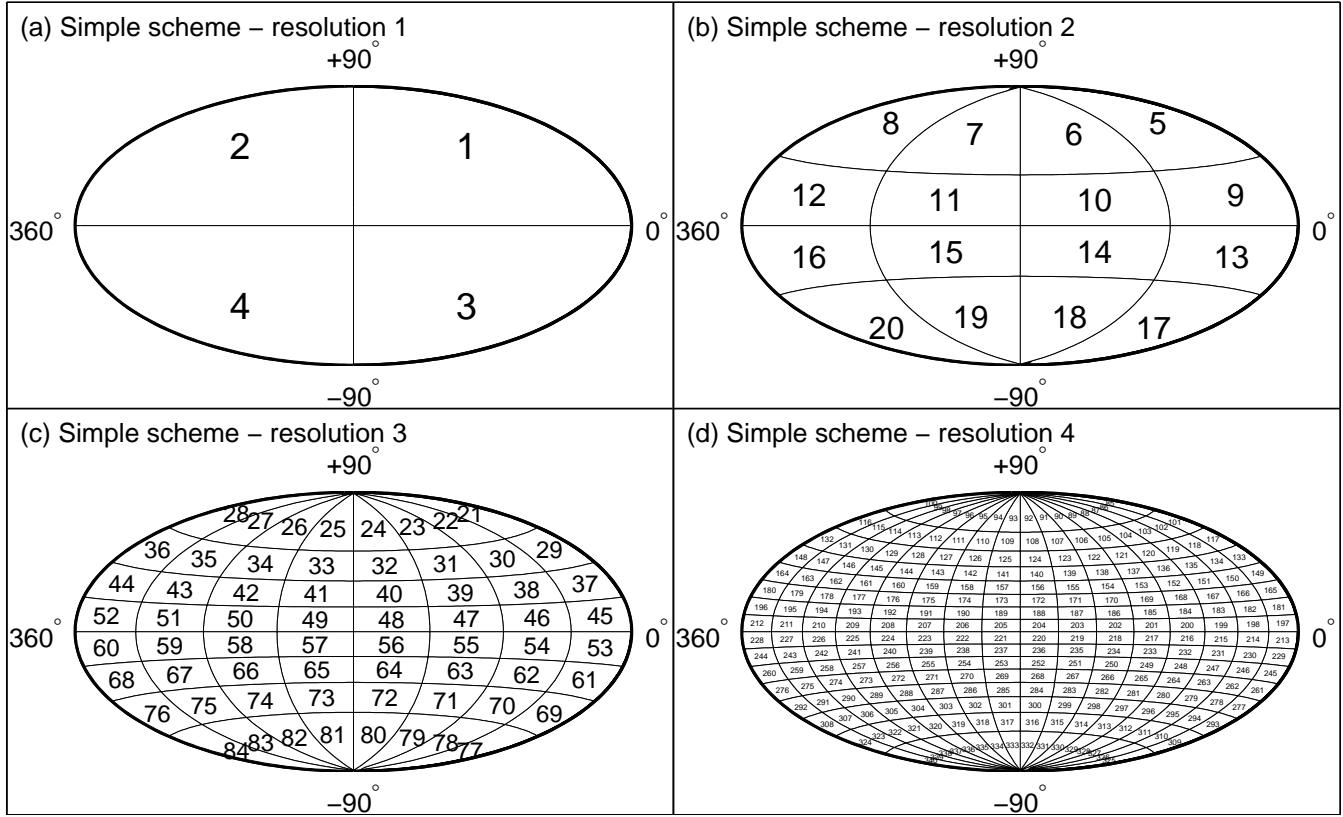}

\caption{\label{fig:simple_pix}The full sky (shown in a Hammer-Aitoff projection
in celestial coordinates) pixelized with the simple pixelization scheme
at the four lowest resolutions.}
\end{figure*}
The underlying concept of pixelization is as follows: before performing
any snapping, balkanization, or unification, divide the mask into
predefined regions called {}``pixels'' and split each polygon along
the pixel boundaries such that each polygon is only in one pixel.
Then for following tasks, polygons need only be compared with other
polygons in the same pixel.

\begin{figure*}
\includegraphics{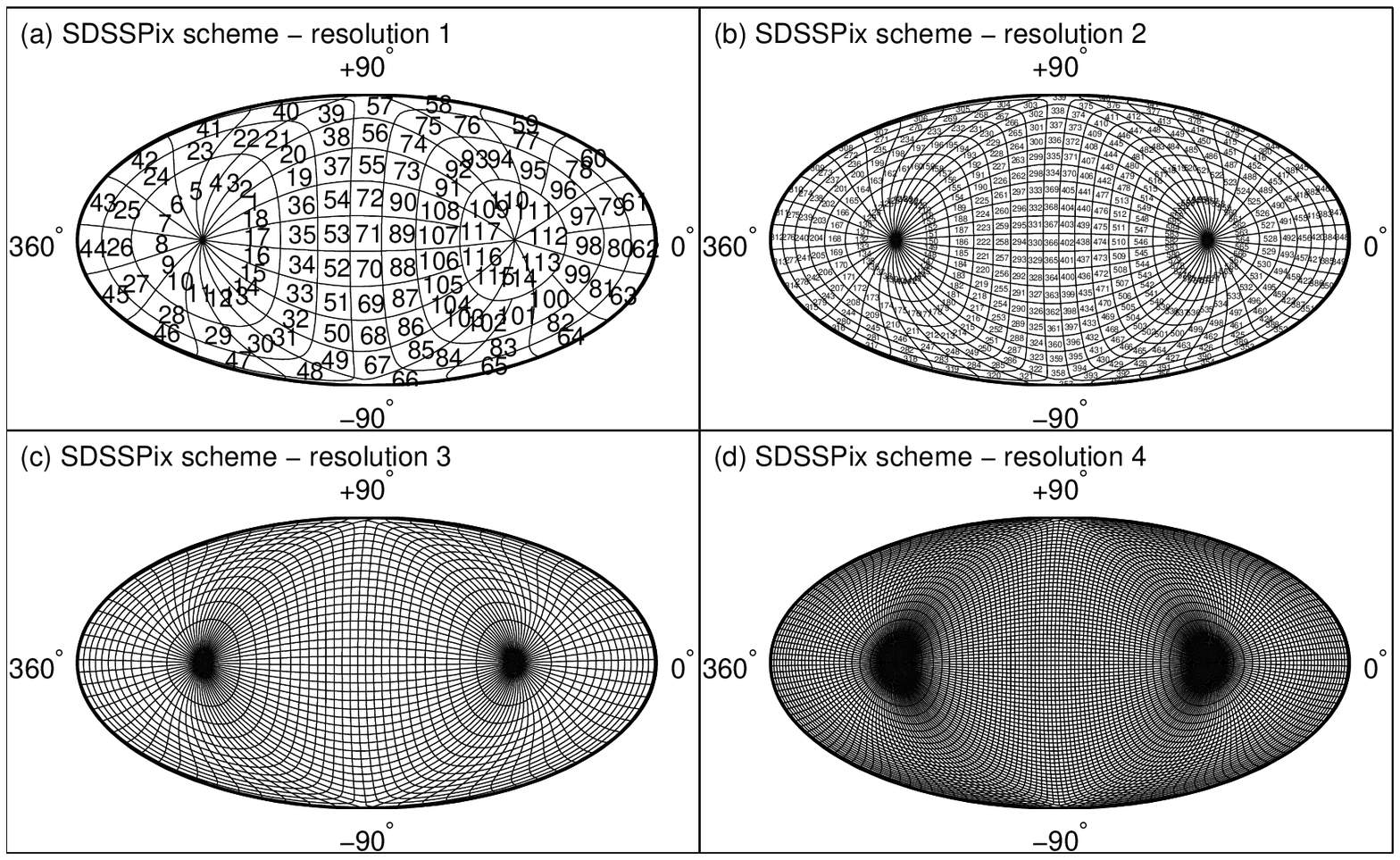}

\caption{\label{fig:sdsspix}The full sky (shown in a Hammer-Aitoff projection
in celestial coordinates) pixelized with the SDSSPix pixelization
scheme at the four lowest resolutions.}
\end{figure*}
The process of balkanization with and without pixelization is illustrated
in Fig.~\ref{fig:pix_concept} for three overlapping polygons. The
top panel shows the unpixelized version: balkanization checks for
overlap between each pair of polygons, and then fragments it into
seven non-overlapping polygons. 

The bottom panel shows the same process but using pixelization as
a first step. First, each polygon is divided along the pixel boundaries,
shown by the dotted lines. At this point, each of the four pixels
shown has three polygons in it: the intersections between that pixel
and each of the original three polygons. Then balkanization is performed
within each pixel: the three polygons in the upper left pixel are
split into five non-overlapping polygons, and so forth, yielding a
final set of 18 non-overlapping polygons. In this illustrative example,
pixelization increases the complexity of the process, but in general
it replaces the $\mathcal{O}\left(N^{2}\right)$ algorithm for balkanization
with one that is roughly $\mathcal{O}\left(M\left(N/M\right)^{2}\right)$
for $N$ polygons and $M$ pixels, which is roughly $\mathcal{O}\left(N\right)$
if $M\sim N$. For large, complicated masks such as SDSS, this speeds
up the processing time by a factor of ${\sim}1200$. 

Once a mask has been pixelized, the important task of determining
in which polygon(s) a given point lies is sped up greatly as well:
one merely has to calculate in which pixel the point lies, and then
test if the point is in each polygon within that pixel. For a typical
pixelization scheme, the appropriate pixel number for a given point
can be found with a simple formula, i.e. an $\mathcal{O}\left(1\right)$
calculation, so pixelization reduces the $\mathcal{O}\left(N\right)$
algorithm of testing every polygon in the mask to $\mathcal{O}\left(N/M\right)$.
This means that with pixelization, this task does not depend on the
total number of polygons in the mask at all if $M\sim N$. 

It is important to note that the pixelization procedure makes no approximations
to the original mask -- the pixels are used simply as a tool to determine
which polygons are close to each other, not as a means of discretizing
the mask itself. A mask that has been balkanized after pixelization
contains all of the same information as it would without pixelization,
except that it took a tiny fraction of the time to produce. Furthermore,
unification can be applied across the whole mask rather than within
each pixel, which effectively unpixelizes the mask if desired. Thus
there are essentially no drawbacks to using our pixelization procedure.

\subsection{Pixelization schemes}

\label{sub:Pixelization-schemes}

\subsubsection{Simple scheme}

The most straightforward means of pixelizing the sky is to use lines
of equal azimuth and elevation as the pixel boundaries. The azimuth
and elevation typically correspond to celestial coordinates -- right
ascension (RA) and declination (Dec) -- in a survey mask. In this
scheme, the whole sky is split into quadrants along the equator and
prime meridian to form the lowest resolution of pixelization. In \noun{mangle}
this is defined as resolution 1. 

To pixelize to higher resolutions, each pixel is split into four child
pixels, with the boundaries at the midpoints of azimuth and of cos(elevation)
within the pixel. This creates pixels with equal area. Thus resolution
2 consists of each of the four resolution 1 quadrants split into four
pixels each, and so forth. This procedure produces a hierarchical
pixel structure known in computer science terminology as a quadtree \citep{markbook}: 
there are $4^{r}$ pixels in this scheme at resolution
$r$, and each of these pixels has four child pixels at resolution
$r+1$ and $r$ parent pixels, one at each lower resolution. Resolution
0 is defined as being the whole sky. 

In \noun{mangle}, the pixels of the simple scheme are numbered as
follows: the whole sky is pixel 0, the four quadrants of resolution
1 are pixels 1, 2, 3, and 4, resolution 2 is pixels 5-20, and so forth.
Thus just one number specifies both the resolution and the pixel location.
At each resolution, the pixels are numbered in a ring pattern, starting
from an elevation of $90^{\circ}$ and along increasing azimuth for
each ring of equal elevation. The simple scheme pixels at the four
lowest resolutions are shown in Fig.~\ref{fig:simple_pix}.

\subsubsection{SDSSPix scheme}

Alternatively, the pixelization can be done such that it is more closely
aligned with the mask of a given survey. In particular, a pixelization
scheme called SDSSPix\footnotemark\footnotetext{\url{http://lahmu.phyast.pitt.edu/~scranton/SDSSPix/}}
has been developed for use with the SDSS geometry. Like the simple
scheme, SDSSPix is a hierarchical, equal-area pixelization scheme,
and it is based on the SDSS survey coordinates $\lambda$ and $\eta$.
As described in \citet{2002AJ....123..485S}, SDSS survey coordinates
form a spherical coordinate system rotated relative to the celestial
coordinate system. The poles are located at $\rmn{RA}=95^{\circ},\rmn{Dec}=0^{\circ}$
and $\rmn{RA}=275^{\circ},\rmn{Dec}=0^{\circ}$ (J2000), which are
strategically located outside the SDSS covered area and in the Galactic
plane. $\eta$ is the azimuthal angle, with lines of constant $\eta$
being great circles perpendicular to the survey equator, and $\lambda$
is the elevation angle, with lines of constant $\lambda$ being small
circles parallel to survey equator. $\lambda=0^{\circ},\eta=0^{\circ}$
is located at $\rmn{RA}=185^{\circ},\rmn{Dec}={32.5}^{\circ}$ with
$\eta$ increasing northward. This configuration has been chosen such
that the stripes produced by the SDSS scanning pattern lie along lines
of constant $\eta$. 

The SDSSPix base resolution is defined by 36 divisions in the $\eta$
direction (equally spaced in $\eta$) and 13 in the $\lambda$ direction
(equally spaced in $\cos\lambda$), for a total of 468 equal-area
pixels. These divisions are chosen such that at a special resolution
level (called the {}``superpixel'' resolution), there is exactly
one pixel across each SDSS stripe. Finally, as in the simple scheme,
higher resolutions are achieved by hierarchically subdividing each
pixel at a given resolution into four smaller pixels. 

SDSSPix has been included in \noun{mangle} by incorporating several
routines from the SDSSPix software package available online.\footnotemark[\value{footnote}]
The numbering scheme in the \noun{mangle} implementation of SDSSPix
differs somewhat from the internal SDSSPix numbering: in \noun{mangle},
the entire sky is pixel 0, as in the simple scheme, but it has 117
child pixels (instead of 4), numbered from 1 to 117. These pixels,
which comprise resolution 1, are not part of the official SDSSPix
scheme, but are created by combining sets of 4 pixels from what is
defined as resolution 2 in \textsc{mangle}. The resolution 2 pixels
are the official SDSSPix base resolution pixels, and are numbered
from 118 to 585. Higher resolutions are constructed through the standard
SDSSPix hierarchical division of each pixel into 4 child pixels. As
in the simple scheme, the pixel number identifies both the resolution
and the pixel position. The superpixel resolution described above
is defined as resolution 5 in \noun{mangle} and contains a total of
7488 pixels.

\noun{Mangle} typically uses RA and Dec as its internal azimuth and
elevation coordinates, so the SDSSPix pixels are constructed as rectangles
in $\eta-\lambda$ coordinates and then rotated into celestial coordinates.
The SDSSPix pixels at the four lowest resolutions are shown in Fig.~\ref{fig:sdsspix}.

\subsubsection{Other schemes}

\label{sub:Other-schemes}

The implementation of pixelization in \noun{mangle} is designed to
be flexible: it is simple for users to add their own scheme as well.
The pixelization uses only four basic routines:

\begin{enumerate}
\item get\_pixel: Given a pixel number, return a polygon representing that
pixel.
\item which\_pixel: Given a point on the sky and a resolution, return the
pixel number containing the point.
\item get\_child\_pixels: Given a pixel number, return the numbers of its
child pixels.
\item get\_parent\_pixels: Given a pixel number, return the numbers of its
parent pixels.
\end{enumerate}
Adding a new pixelization scheme simply requires creating appropriate
versions of these four routines. Note that it also requires that the
pixels can be represented as polygons -- this is \emph{not} strictly
the case for the \emph{HEALPix} pixels, as discussed further in \S\ref{sec:Unification-with-HEALPix}.

\subsection{Pixelization algorithm}

\label{sub:Pixelization-algorithm}

\begin{figure*}
\includegraphics{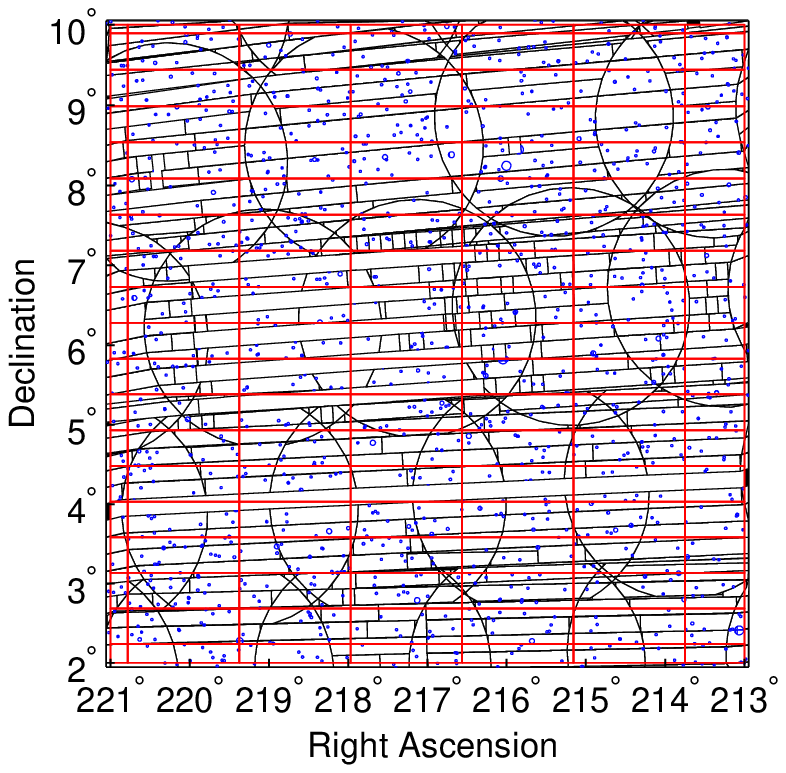}\includegraphics{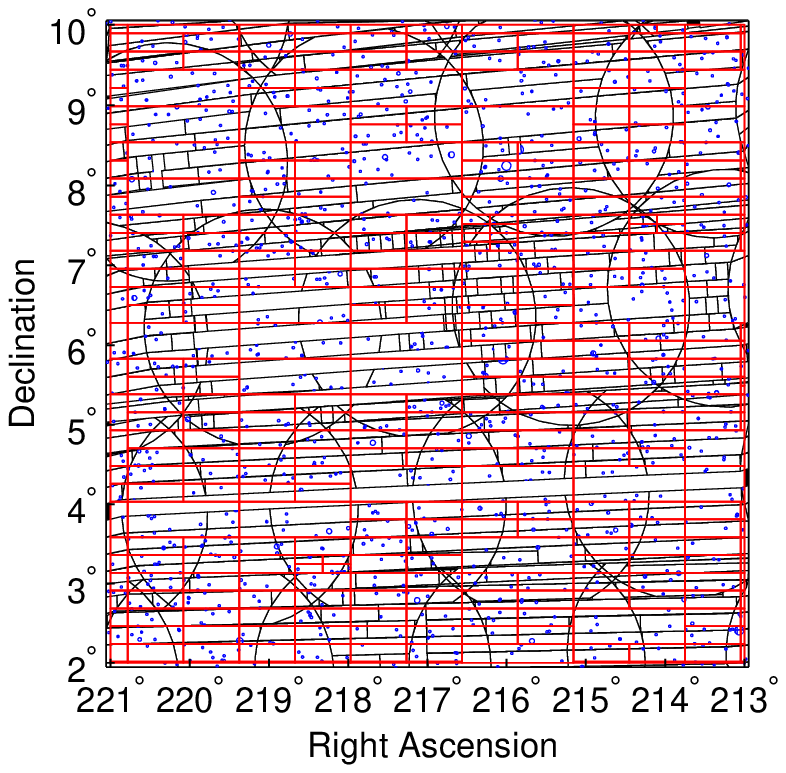}

\caption{\label{fig:examplemask_pixelized}Left: portion of SDSS mask from
Fig.~\ref{fig:examplemask} pixelized to resolution 8 ($4^{8}$ total
pixels on the sky) with the simple pixelization scheme. Pixel boundaries
are shown in red/light gray. Right: the same mask pixelized with the
simple pixelization scheme using the adaptive resolution method with
a maximum of 30 polygons per pixel.}
\end{figure*}
The purpose of pixelization is to speed up the processing of angular
masks, which means that the pixelization itself must be done with
a clever, speedy algorithm or nothing will be gained. The naive algorithm
is to search through all of the polygons in the mask for those that
overlap that pixel. This is $\mathcal{O}\left(NM\right)$ for $N$
polygons and $M$ pixels, and is not sufficient for our purposes.

Our fast pixelization algorithm is a recursive method that takes advantage
of the hierarchical nature of the pixelization schemes. The method
works as follows:

\begin{enumerate}
\item Start with all the mask polygons that are in pixel $i$.
\item Create polygons for each child pixel of pixel $i$ at the next resolution
level.
\item Split the mask polygons in pixel $i$ along the child pixel boundaries
such that each polygon lies within one child pixel.
\item Repeat steps 1-3 for the mask polygons in each child pixel until desired
stopping point is reached.
\end{enumerate}
Starting this with pixel $i=0$, i.e., the whole sky, will pixelize
the entire mask in $\mathcal{O}\left(N\log M\right)$ time. Thus the
pixelization does not add too much overhead time to the overall mask
processing.

There are two different methods for choosing the desired stopping
point of the pixelization. The simplest method is to stop at a fixed
resolution, such that the entire mask is pixelized with pixels of
the same size. The example mask from Fig.~\ref{fig:examplemask}
is shown in the left panel of Fig.~\ref{fig:examplemask_pixelized}
pixelized to a fixed resolution in the simple scheme.

Alternatively, the stopping condition can be chosen to be a maximum
number of polygons allowed in each pixel: if there are more than $N_{\rmn{max}}$
polygons in a given pixel, continue the recursion and divide those
polygons into pixels at the next resolution level. This results in
an adaptively pixelized mask, where higher resolutions are automatically
used in regions of the mask that are more complicated. This method
is especially useful for masks with varying degrees of complexity
in different areas. The example mask from Fig.~\ref{fig:examplemask}
is shown again in the right panel of Fig.~\ref{fig:examplemask_pixelized}
pixelized adaptively with $N_{\rmn{max}}=30$.

The implementation of pixelization in \noun{mangle} allows users to
choose either of these methods and to select values for the fixed
resolution level or for $N_{\rmn{max}}$.

\subsection{Speed trials}

\begin{figure*}
\includegraphics{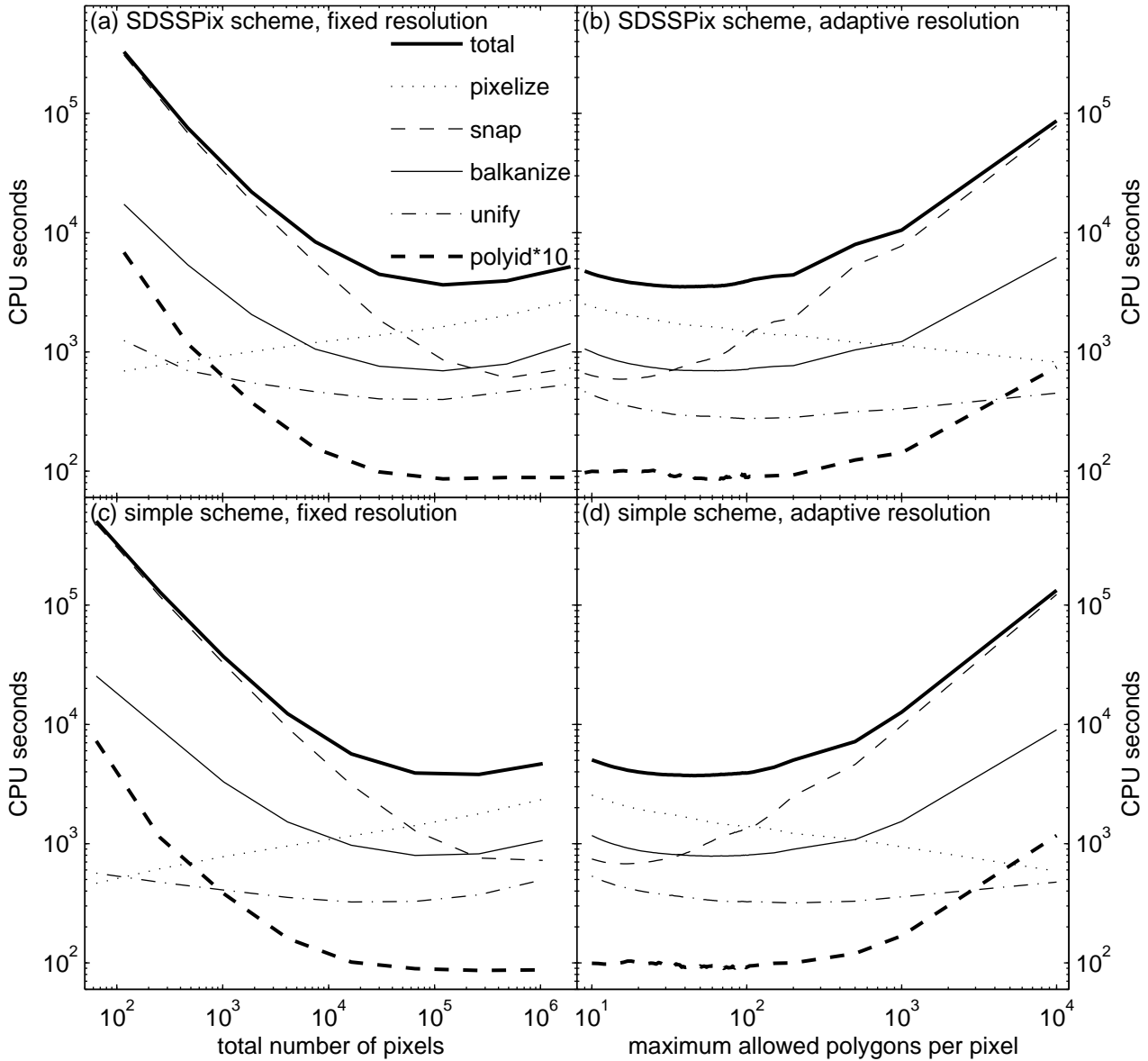}

\caption{\label{fig:speedtrials}Time required to process the full SDSS DR5
mask with different choices of pixelization schemes and methods. The
{}``total'' curves are the sum of the pixelization, snapping, balkanization,
and unification curves. Also shown is the time required to identify
in which polygon each of the ${\sim}400{,}000$ SDSS DR5 galaxies
lies (polyid), scaled up by a factor of 10. The plots on the left
show time vs. the number of pixels at a fixed resolution. The plots
on the right show time vs. $N_{\rmn{max}}$, the maximum number of
polygons in allowed in each pixel when pixelizing with the adaptive
resolution method.}
\end{figure*}
In order to choose optimal default values for the maximum resolution
and $N_{\rmn{max}}$ as well as to demonstrate the dramatic improvements
in speed that pixelization provides, we have conducted a series of
speed trials of the new \noun{mangle} software.

There are two procedures we are interested in optimizing: firstly,
basic processing of a mask detailed in \S\ref{sec:Mangle-terminology}
involving pixelization, snapping, balkanization, and unification,
and secondly, the use of the final mask to identify in which polygon
a given point lies. We carried out trials of these two procedures
using both the simple and SDSSPix pixelization schemes described in
\S\ref{sub:Pixelization-schemes}. For each of these schemes, we
tested both the fixed and adaptive resolution methods for stopping
the pixelization algorithm described in \S\ref{sub:Pixelization-algorithm}.
For the fixed resolution method, we measured the time for several
different values of the maximum resolution, and for the adaptive resolution
method, we measured the time as a function of $N_{\rmn{max}}$. 

The results are shown in Fig.~\ref{fig:speedtrials}. (Note that
the overhead time for reading and writing files and doing general
setup has been subtracted -- the times shown here are just for the
primary operations.) From the fixed resolution trials, we see that
the optimal resolution choice for the SDSS mask is the one that has
approximately $10^{5}$ total pixels on the sky for both the simple
and SDSSPix schemes. This corresponds to resolution 9 for the simple
scheme and resolution 6 for SDSSPix. When using the adaptive method,
the choice for the maximum number of polygons allowed in each pixel
that gives the fastest processing is $N_{\rmn{max}}=40$ for the simple
scheme and $N_{\rmn{max}}=46$ for SDSSPix. Overall, the fastest choice
(by a slight margin) for the SDSS mask is using the SDSSPix scheme
with adaptive resolution. It is also interesting to note that different
\noun{mangle} processes have different optimum values -- for example,
snapping is fastest when there are fewer polygons in each pixel compared
to balkanization.

The impact of our pixelization algorithm is most clearly demonstrated
by Fig.~\ref{fig:speed_fractions}. This shows the processing time
and polygon identification time for a series of selected portions
of the SDSS mask as a function of the total number of polygons, both
with and without pixelization -- again, overhead time has been subtracted
here. Pixelization clearly gives an improvement in speed that becomes
increasingly significant for larger numbers of polygons. 

To quantify this, we fit theoretical models to each series of trials.
Without pixelization, the processing time for snapping, balkanization,
and unification is well-fitted by $AN^{2}$ where $A$ is a free parameter
with a best-fitted value of $A=3.2\times10^{-5}$ CPU seconds. With
pixelization (and using the adaptive resolution method with the optimal
choice for $N_{\rmn{max}}$) this is reduced substantially. We fit
the times using theoretical models based on the formulas in \S\ref{sub:Pixelization-concept}
assuming $M\sim N$. We fit the pixelization time with $BN\log_{4}N$
and the snapping, balkanization, and unification time with $CN$ where
$B$ and $C$ are free parameters. The simple scheme gives $\left(B,\, C\right)=\left(0.0054,\,4.7\times10^{-4}\right)$
and the SDSSPix scheme gives $\left(B,\, C\right)=\left(0.0045,\,4.8\times10^{-4}\right)$
(units are all in CPU seconds) -- thus for processing the SDSS mask,
the SDSSPix scheme is slightly faster than the simple scheme. Our
fits give $C/B=0.1$, so the overall processing time scales like $\mathcal{O}\left(N+0.1N\log_{4}N\right)$. 

These fitted curves are shown in Fig.~\ref{fig:speed_fractions}
and allow us to extrapolate estimates for processing masks with larger
numbers of polygons. For the full SDSS DR5 mask containing about 300,000
polygons, pixelization reduces the processing time by a factor of
${\sim}1200$. The improvement for future surveys will be even more
dramatic: the mask for the co-added LSST survey might contain ${\sim}10^8$
polygons -- pixelization reduces the processing time by a factor of
${\sim}24{,}000$.

The time for identifying in which polygon each SDSS galaxy lies is
significantly sped up as well: without pixelization it is fitted by
$DN$ where $D=0.012$ CPU seconds, although the scatter is rather
large due to dependence on which polygons are used. With pixelization,
it is well-fitted by a constant (8.2 CPU seconds for the simple scheme,
7.9 CPU seconds for the SDSSPix scheme) -- thus the time for polygon
identification does not depend on how many polygons are in the mask
if the number of pixels used is chosen to be roughly proportional to the 
number of polygons.
For the SDSS mask, the time to identify the polygons for all of the
SDSS galaxies is reduced by a factor of nearly 500.

\section{Unification with \emph{HEALPix} and other pixelized tools}

\label{sec:Unification-with-HEALPix}

\begin{figure*}
\includegraphics{fig2b}\includegraphics{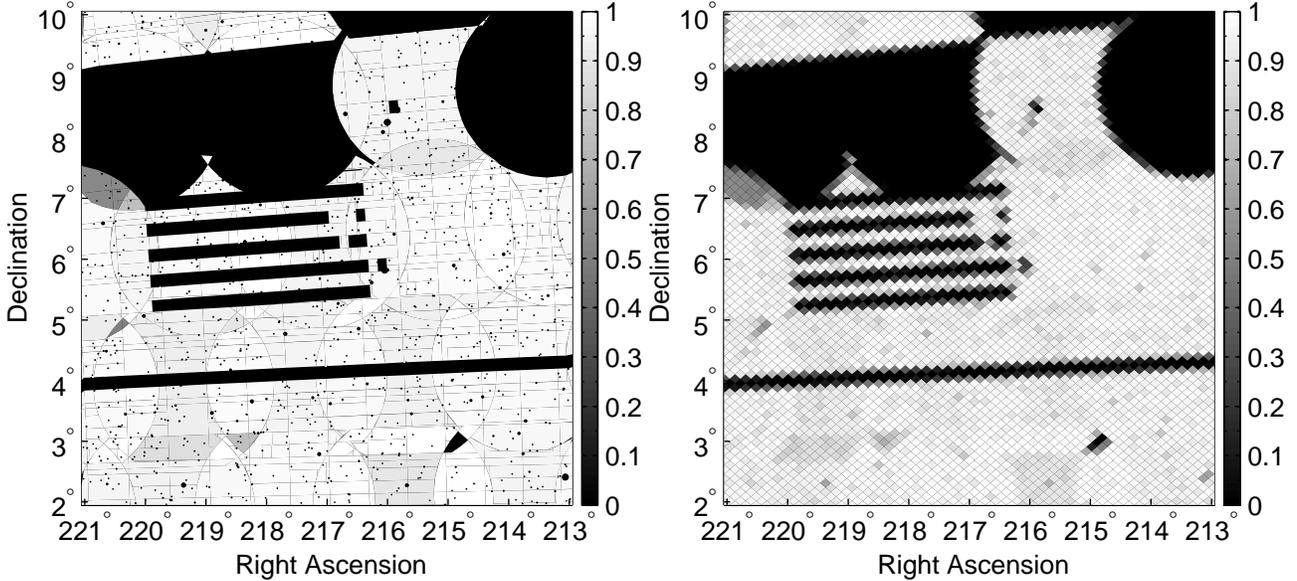}

\caption{\label{fig:rasterized}Left: Portion of SDSS mask as shown in Fig.~\ref{fig:examplemask}.
Right: Portion of SDSS mask from Fig.~\ref{fig:examplemask} as approximated
by \emph{HEALPix} pixels, rasterized to $N_{side}=512$.}
\end{figure*}
\begin{figure*}
\includegraphics[width=12cm]{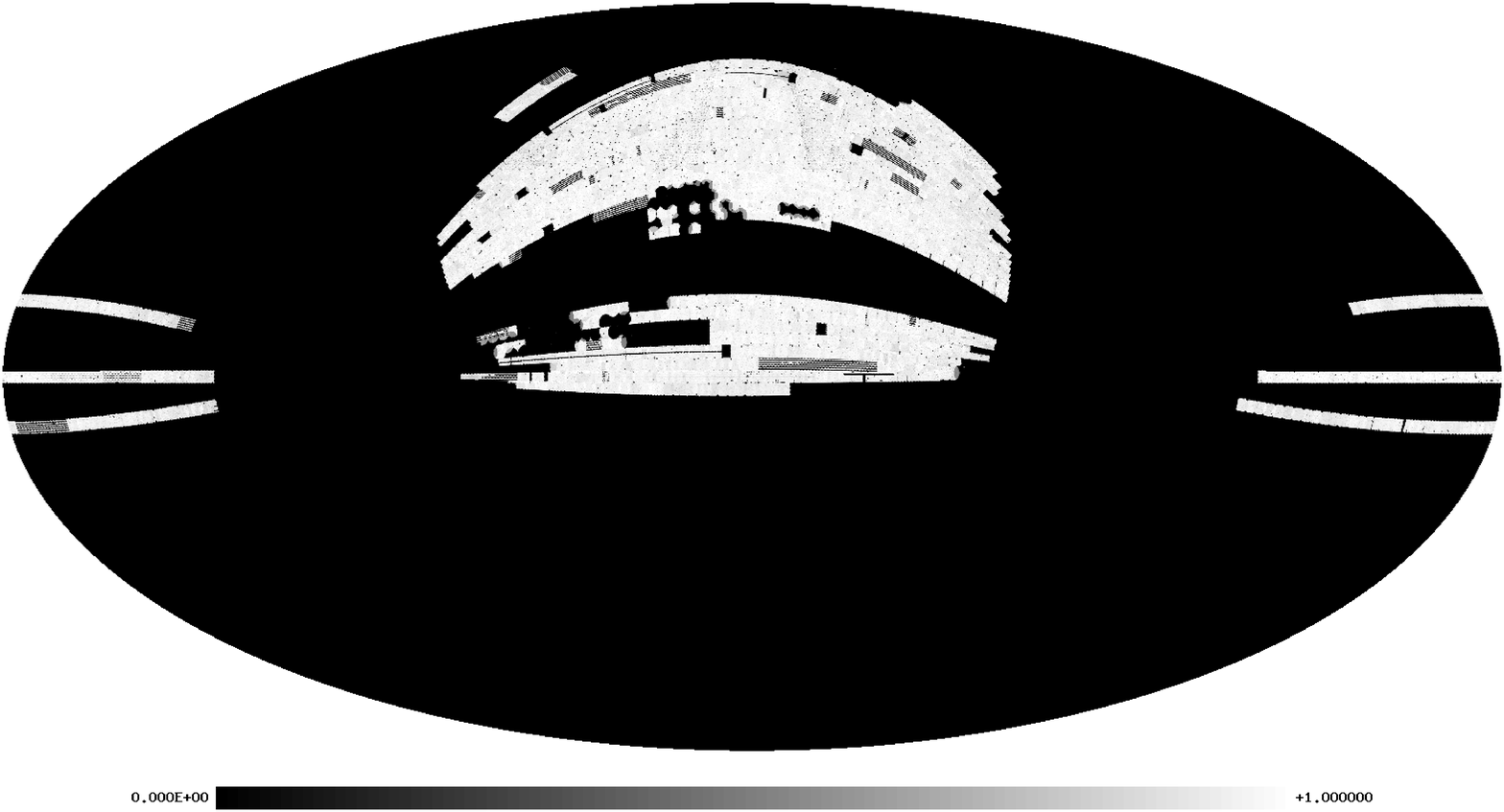}

\includegraphics[width=12cm]{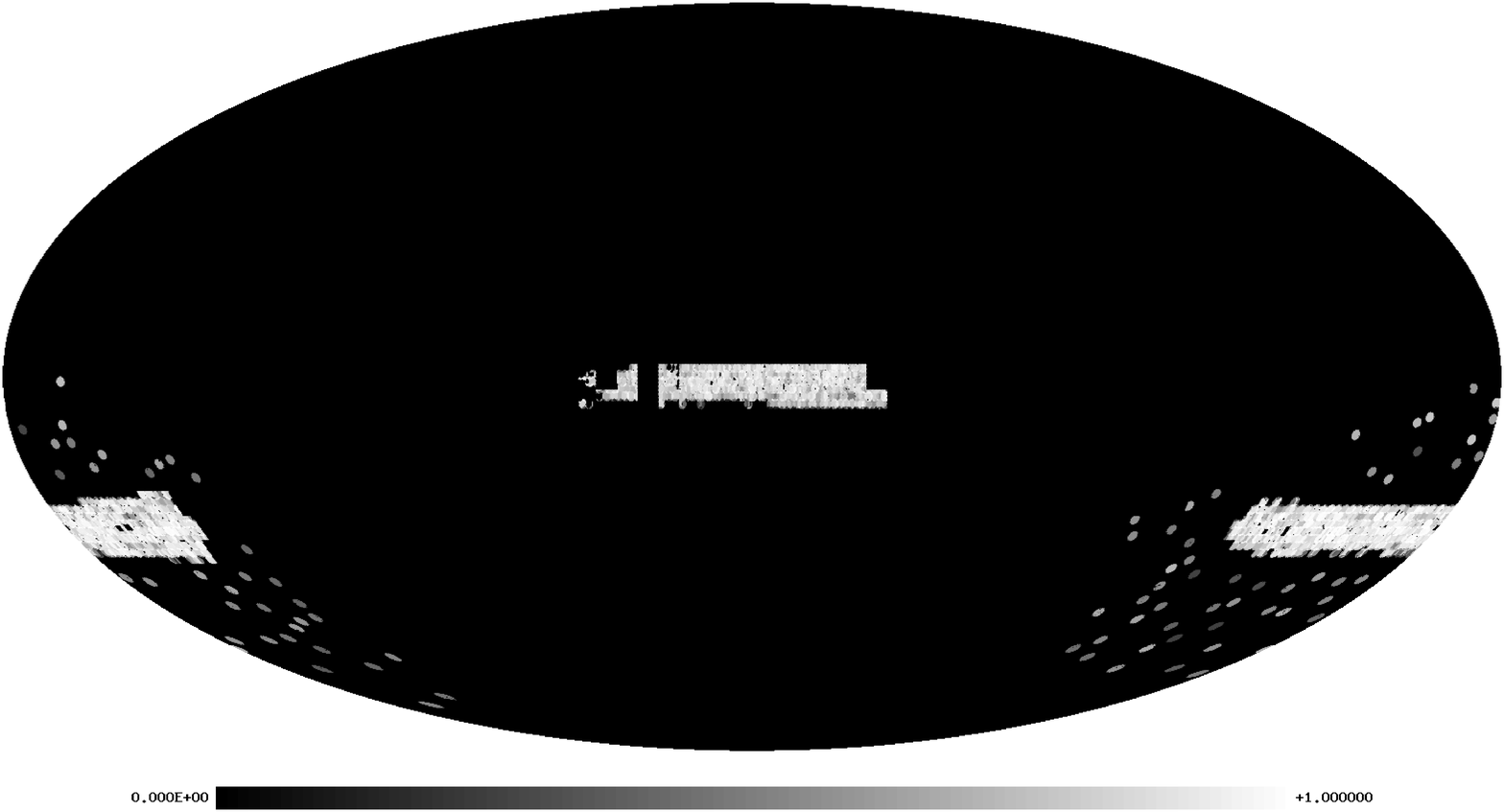}

\includegraphics[width=12cm]{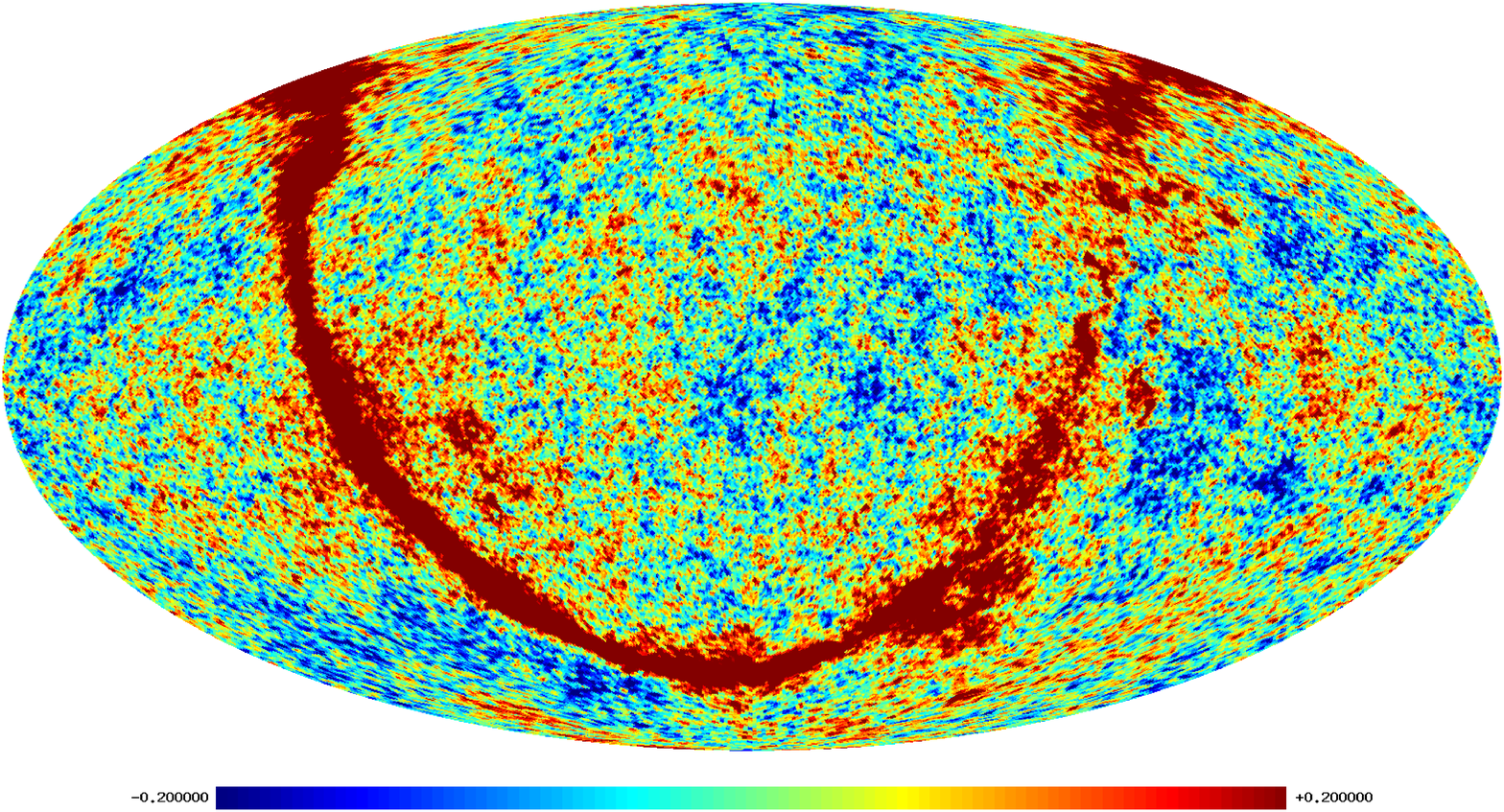}

\caption{\label{fig:masksandcmb}The new version of \noun{mangle} can output
angular masks in \emph{HEALPix} format, which allows for easy comparisons
to CMB and other sky map data and allows users to take advantage of
existing \emph{HEALPix} tools. Top: The SDSS DR5 completeness mask,
rasterized and plotted using HEALPix routines. Middle: The final 2dFGRS
completeness mask as determined in \citet{Hamilton_in_prep}, rasterized
and plotted using \emph{HEALPix} routines. Bottom: CMB temperature
difference map measured by WMAP channel 4, with units in mK. \citet{2007ApJS..170..377S}
All three maps are shown at $N_{side}=512$ in celestial coordinates.}
\end{figure*}
\emph{HEALPix} (G{\'o}rski et al. 1999b, 2005)\nocite{2005ApJ...622..759G,1999astro.ph..5275G}
is a hierarchical, equal-area, isolatitudinal pixelization scheme for
the sphere motivated by computational challenges in analyzing CMB
data (G{\'o}rski et al. 1999a, \citealt{1999CoScE...1...21B}). \nocite{1999elss.conf...37G} Its base resolution
consists of 12 equal-area pixels; to generate higher resolutions,
each pixel at a given resolution is hierarchically subdivided into
four smaller pixels, and it represents an interesting class of spherical
projections \citep{2007MNRAS.381..865C}. \emph{}Resolution in \emph{HEALPix}
is defined in terms of $N_{\rmn{side}}$, the number of divisions
along the side of a base-resolution pixel required to reach the desired
resolution. Because of the hierarchical definition of the higher resolution
pixels, $N_{\rmn{side}}$ is always a power of 2, and the total number
of pixels at a given resolution is $12N_{\rmn{side}}^{2}$.

\emph{HEALPix} is a very useful scheme in that it allows for fast
and accurate astrophysical computations by means of appropriately
discretizing functions on the sphere to high resolution (\citealt{1998astro.ph..3317W}, Dor{\'e} et al. 2001b).\nocite{2001A&A...374..358D}
In particular, \emph{HEALPix} includes routines for fast computations
of spherical harmonics (Dor{\'e}, Knox, \& Peel 2001a; \citealt{2002ApJ...567....2H,2001PhRvD..64h3003W,2001ApJ...561L..11S}). \nocite{2001PhRvD..64h3001D}
It is widely used in the analysis of cosmic microwave background data
from WMAP \citep{2007ApJS..170..377S} and has recently been used
to approximate galaxy survey masks as well \citep{2007ApJ...657..645P}. 

Combining \emph{HEALPix} with \noun{mangle} is useful because it facilitates
comparisons between galaxy survey data and the CMB as is done in experiments
measuring the integrated Sachs-Wolfe effect \citep{2005PhRvD..72d3525P,2006PhRvD..74f3520G,2007MNRAS.377.1085R},
the Sunyaev-Zel'dovich effect \citep{2004MNRAS.347L..67M,2006ApJ...651..643R},
etc., and can also be used for generating masks that block out regions
of high dust extinction from galaxy surveys (\citealt{1998ApJ...500..525S},
dust map available in \emph{HEALPix} format online%
\footnote{\protect\url{http://lambda.gsfc.nasa.gov/product/foreground/ebv_map.cfm} %
}).

In general, it can be applied to any task requiring comparison between
a piecewise-constant function on the sphere to a continuous function
sampled on a discretized spherical grid. Converting an angular mask
into \emph{HEALPix} format also allows for rapid computations of approximate
spherical harmonics of the mask using the existing \emph{HEALPix}
tools. 

The implementation of the \emph{HEALPix} scheme in \textsc{mangle}
consists of two components:

\begin{enumerate}
\item A new polygon format, {}``healpix\_weight'', which allows the user
to input a list of weights corresponding to each \emph{HEALPix} pixel
at a given $N_{\rmn{side}}$ parameter; 
\item A new utility, {}``rasterize'', which essentially allows the user
to pixelize a mask against the \emph{HEALPix} pixels, by means of
a different technique than the pixelization method described in \S\ref{sec:Speedup:-pixelization}. 
\end{enumerate}
Together, these new features allow effective two-way conversion between
the \emph{HEALPix} specifications and those of \textsc{mangle}.

\subsection{Importing \emph{HEALPix} maps into \noun{mangle}}

\label{sub:Importing-HEALPix-files}

The structure of the healpix\_weight format is quite simple: an input
file consists of a list of numbers corresponding to the weight of
each \emph{HEALPix} pixel at a given $N_{\rmn{side}}$ parameter,
using the nested numbering scheme described in \citet{2005ApJ...622..759G}.
In addition, the definition of the $N_{\rmn{side}}$ parameter is
extended to include 0, which corresponds to a single pixel covering
the entire sphere. \textsc{mangle} constructs polygons approximately
equivalent to the \emph{HEALPix} pixels through the following procedure:

\begin{enumerate}
\item The exact azimuth and elevation of each vertex of a given pixel are
calculated using the \emph{HEALPix} utility {}``pix2vec\_nest''; 
\item The exact azimuth and elevation of each vertex of the four child pixels
of the current pixel are calculated using the same utility, four of
which are the midpoints of the edges of the current pixel; 
\item The four vertices of the given pixel are combined with the four midpoints
to construct the current pixel. Each edge is defined by the circle
that passes through the two vertices and the midpoint, using the {}``edges''
format described in \citet{2004MNRAS.349..115H}.
\item To eliminate spurious antipodal pieces of the polygon defining the
pixel, a fifth cap is constructed whose axis coordinates are the exact
centre of the current pixel and whose radius is $10^{-6}$ radians
greater than the distance from the centre of the pixel to any of the
four vertices---this cap thus encloses the entire pixel. 
\end{enumerate}
A similar technique could be applied to incorporate other pixelization
schemes not exactly described by spherical polygons as well. It is
important to note that the pixels constructed in this manner are not
exactly equivalent to the actual \emph{HEALPix} pixels -- while the
hierarchical and isolatitudinal properties of the \emph{HEALPix} pixels
are preserved, the equal area property is not. For example, at $N_{\rmn{side}}=1$,
the areas of the approximate pixels differ on average from the actual
area by about $0.08\%$. 

However, as the $N_{\rmn{side}}$ parameter increases, this difference
decreases rapidly: at $N_{\rmn{side}}=512$, the average difference
from the actual area is $0.000002\%$. The boundaries of the both
the actual \emph{HEALPix} pixels and our circles approximating them
become straight lines in the flat-sky approximation, i.e., when the
pixel size is much less than 1 radian. Thus for the resolutions at
which \emph{HEALPix} is typically used, the difference is totally
negligible. 

In applications involving integrations over the sphere, such as calculating
spherical harmonics, the slight area differences between the pixels
can be corrected for in a straightforward manner since the area of
each pixel is known (and can be easily extracted using \noun{mangle}'s
{}``area'' format): simply multiply the value of the function in
each pixel by the area of that pixel divided by the average pixel
area, and then the \emph{HEALPix} spherical harmonics routines will
be exact. Furthermore, the paranoid user can get higher precision
by rasterizing to a higher resolution and then using the {}``ud\_grade''
\emph{HEALPix} utility to obtain results for lower resolutions.

\subsection{Exporting polygon files as \emph{HEALPix} maps}

The idea behind the rasterization method used to convert polygon files
into \emph{HEALPix} maps is very similar to that of pixelization:
to split up the polygons that comprise a mask using a given set of
pixels, such that afterward each polygon lies in only one pixel. However,
rasterization is somewhat different, in that afterward the converse
statement also holds: each pixel contains only one polygon, namely,
itself. In particular, rasterization uses an arbitrary user-defined
spherical pixelization as the pre-determined scheme against which
to split up the polygons in a given input mask. In general, the user-defined
{}``rasterizer'' pixels may be from any pixelization of the sphere,
but the method was originally developed for use with the approximate
\emph{HEALPix} pixels described in the previous section. From a practical
standpoint, it would be simplest to implement any spherical pixelization
that can be exactly represented as spherical polygons by following
the steps in \S\ref{sub:Other-schemes}; however, for pixelization
schemes that do not possess this property (such as \emph{HEALPix}),
rasterization provides an alternate method of implementation.

The final product of rasterization is a polygon file in which each polygon
corresponds to one of the rasterizer pixels (either the approximate
\emph{HEALPix} pixels or a user-defined set of pixels) and its weight
is the area-averaged weight of the input angular mask within that
pixel. This output can easily be converted into a FITS file read by
the \emph{HEALPix} software using a simple script provided with \noun{mangle}.

Rasterization consists of the following steps:

\begin{enumerate}
\item Calculate the area of each rasterizer pixel;
\item Compute the area of the intersection of each input mask polygon with
each rasterizer (e.g., \emph{HEALPix}) pixel;
\item Calculate the area-averaged weight within each rasterizer pixel. 
\end{enumerate}
Note that, as with snapping, balkanization, and unification, this
procedure is greatly accelerated by pixelizing both the rasterizer
polygons and the polygons defining the mask to the same resolution
using one of the pixelization schemes described in \S\ref{sub:Pixelization-schemes},
e.g. simple or SDSSPix. Then step (ii) in the above procedure then
involves comparing only polygons in the same simple/SDSSPix pixel.
Our example mask from Fig.~\ref{fig:examplemask} (duplicated in
the left panel of Fig.~ \ref{fig:rasterized}) is shown in the right
panel of Fig.~\ref{fig:rasterized} rasterized with \emph{HEALPix}
pixels at $N_{\rmn{side}}=512$. The ability to convert between \noun{mangle}
and \emph{HEALPix} formats allows for straightforward comparisons
between different types of functions defined on the sphere, e.g. angular
masks of galaxy surveys and the CMB, as illustrated in Fig.~\ref{fig:masksandcmb}.

\section{Summary}

\label{sec:Summary}

As technologies for surveying the sky continue to improve, the process
of managing the angular masks of galaxy surveys grows ever more complicated.
The primary purpose of this paper has been to present a set of dramatically
faster algorithms for completing these tasks. 

These algorithms are based on dividing the sky into regions called
{}``pixels'' and performing key operations only within each pixel
rather than across the entire sky. The pixelization is based on a hierarchical
subdivision of the sky -- this produces a quadtree data structure that keeps
track of which polygons are nearby each other. The preprocessing step of
pixelization is $\mathcal{O}\left(N\log N\right)$ for a mask with $N$ polygons
and ${\sim}N$ pixels,
and it reduces
the mask processing time from $\mathcal{O}\left(N^{2}\right)$
to $\mathcal{O}\left(N\right)$.
Furthermore, it reduces
the time required to locate a point within a polygon from $\mathcal{O}\left(N\right)$
to $\mathcal{O}\left(1\right)$.
 
This method is exact, i.e., it
does not make a discrete approximation to the mask, 
and it takes only a tiny fraction of the computation time.
It accelerates the processing of the SDSS mask
by a factor of about 1200 and 
reduces the time to locate all of the
galaxies in the SDSS mask by a factor of nearly 500.
It will provide even more dramatic gains
for future surveys: processing time for the LSST large-scale structure
mask could be reduced by a factor of about 24,000.

We have also described a method for converting between masks described
by spherical polygons and sky maps in the \emph{HEALPix} format commonly
used by CMB and large-scale structure experiments. This provides a
convenient way to work with both piecewise-constant functions on the
sky such as the completeness of a galaxy survey and continuously varying
sky maps such as the CMB temperature. Converting angular masks into \emph{HEALPix}
format also allows users to take advantage of existing \emph{HEALPix} tools
for rapidly computing spherical harmonics.

All of the new algorithms and features detailed here have been integrated
into the \noun{mangle} software suite, which is available for free
download at \texttt{\url{http://space.mit.edu/home/tegmark/mangle/}}.
This updated software package should prove increasingly useful in the
coming years, especially as next-generation surveys such as DES, WFMOS,
Pan-STARRS, and LSST get underway.

\section*{Acknowledgments}

We thank Michael Blanton for providing the SDSS DR6 VAGC mask, Krzysztof
G\'{o}rski and collaborators for creating the \emph{HEALPix} package
(G{\'o}rski et al. 1999b, 2005),\nocite{2005ApJ...622..759G,1999astro.ph..5275G}
and the WMAP team
for making their data public via the Legacy Archive for Microwave
Background Data Analysis (LAMBDA)%
\footnote{The WMAP \protect\url{http://lambda.gsfc.nasa.gov}. %
}. Support for LAMBDA is provided by the NASA Office of Space Science.
This work was supported by NASA grant NNG06GC55G, NSF grants AST-0134999,
0607597 and 0708534, the Kavli Foundation, and fellowships from the
David and Lucile Packard Foundation and the Research Corporation.

\bibliographystyle{mn2e}
\bibliography{mangle,mangle1}

\end{document}